\pdfoutput=1
\documentclass[conference]{IEEEtran}
\usepackage{graphicx}
\usepackage{balance}
\usepackage{paralist}
\usepackage{color}
\usepackage{amsmath}
\usepackage[ruled]{algorithm2e}

\usepackage[labelfont=bf]{caption}
\usepackage{subcaption}
\usepackage[font=scriptsize]{subcaption}
\usepackage{float}

\begin{document}

\title{Discover Aggregates Exceptions over Hidden Web Databases}

\author{
    \IEEEauthorblockN{
        Saad Bin Suhaim\IEEEauthorrefmark{1},
        Weimo Liu\IEEEauthorrefmark{1},
        Nan Zhang\IEEEauthorrefmark{1}
    }
    \IEEEauthorblockA{
       \IEEEauthorrefmark{1}The George Washington University\\
        \{ssuhaim, wliu, nzhang10\}@gwu.edu
    }
}

\maketitle

\begin{abstract}
Nowadays, many web databases ``hidden" behind their restrictive search interfaces (e.g., Amazon, eBay) contain rich and valuable information that is of significant interests to various third parties. Recent studies have demonstrated the possibility of estimating/tracking certain aggregate queries over dynamic hidden web databases. Nonetheless, tracking all possible aggregate query answers to report interesting findings (i.e., exceptions), while still adhering to the stringent query-count limitations enforced by many hidden web databases providers, is very challenging.  In this paper, we develop a novel technique for tracking and discovering exceptions (in terms of sudden changes of aggregates) over dynamic hidden web databases. Extensive real-world experiments demonstrate the superiority of our proposed algorithms over baseline solutions.
\end{abstract}

\section{Introduction}\label{intro}
In recent years, decision support and data mining systems have became very important for analyzing very large databases. Yet these systems have limited functionality and lack of some important features. One of the very important features that these systems lack is the ability to detect and report  interesting finding (e.g., exceptions). Moreover, these systems cannot directly be applied on hidden web databases in a time where these databases could have a rich and valuable content that interests various third parties.\par
In this paper, we develop a novel technique for discovering $exceptions$ of aggregate queries, e.g., AVG, SUM, over $dynamic$ hidden web databases.\par
\noindent\textbf{Change Detection:} Monitor and report changes in databases is very important for decision support and data mining systems. These systems benefit from executing aggregate queries (e.g., SUM, AVG) on very large databases to detect and report interesting finding, which can often be very expensive and resource intensive. Many systems use pre-computation of aggregates to improve response time. Although they do not calculate all aggregates (seen as view selection problem) and relay on a predetermined number of queries. An examples of such systems is the online analytical processing (OLAP).\par
\noindent\textbf{Hidden Web Databases:} The idea of hidden web databases is that they are $hidden$ behind restrictive search interfaces. These databases are not accessible through the traditional search engines, and the only way to access them is by submitting desired values for one or more attributes (to form a conjunctive search query) and get a small number (bounded by a constant $k$ which can be 50 or 100) of tuples that match the specified query. This type of databases is very common on the Internet, and some examples of such databases include Amazon.com, Yahoo! Autos, eBay.com, etc.\par
\noindent\textbf{Problem Motivation:} Hidden web databases contain a rich content that interests various third parties, such as government agencies, and academic and commercial sectors. These parties could benefit from the ability to monitor a wide variety of aggregate queries and report interesting finding, since these aggregates are the most common type of queries in decision support systems. For example, durring Black Friday/Cyber Monday of 2014, Amazon announced a huge discount on their new unlocked Amazon Fire smartphone from \$499 to \$199 (a \$250 price drop). This announcement was resulted by a sharp drop in the average price of the same smartphone on the other websites (i.e., the average price on eBay for all Amazon Fire smartphones droped from \$310 to \$258). More generally, when the AVG price for a certain product that listed on a sale database increases rapidly, this may indicate an increase in demand for this product. Similarly, when the AVG salary offered on job database that require a certain skill  increases quickly,  this  may  indicate  an  expansion  of  the corresponding market.\par
\noindent\textbf{Challenges:} Discovering exceptions of aggregate queries on hidden web databases has two critical challenges: \begin{inparaenum}[i\upshape)] \item the challenge of hidden web databases in general; and \item the challenge of dynamic aggregate estimation.\end{inparaenum}
\begin{enumerate}[i\upshape)] \item \label{itm:challenge of hidden web databases} Challenge of hidden web databases: Most of real-world hidden web databases do not directly support aggregate queries through their web interfaces. The only way to answer such aggregates is by combining multiple search queries. A problem with this solution is that most real-world web databases apply limitation to the number of search queries one can issue through per-IP and per-developer (i.e., eBay limits API calls to 5,000 per day). Prior work \cite{static aggregate} discussed such way to estimate aggregate queries for static databases (i.e., assuming databases do not change overtime). This is unreasonable assumption, since in reality, most real-world web databases are frequently updated. An improved technique has been introduced by \cite{dynamic aggregate} to solve the limitation problem over dynamic databases. However, their solution only consider one single aggregate at a time. To discover exceptions over a dynamic database, we need to monitor a large group of aggregates, which make solving the limitation problem more challenging.

\item Challenge of dynamic aggregate estimation: Prior work (e.g., \cite{dynamic aggregate}) has introduced techniques to overcome (\ref{itm:challenge of hidden web databases}) up to a certain level and estimates aggregates over dynamic hidden web databases. Their existing algorithm applies a random walk technique and distributes the query budget available for reissuing (i.e., updating) previous drill downs, and the rest for initiating new drill downs to track and estimate various types of aggregates. This technique wastes a lot of queries by producing independent samples, such that each sample can only be used to estimate one candidate once (e.g., no information is retained/reused to other aggregate queries).\end{enumerate}\par
In contrast, applying traditional decision support systems to this problem requires a fully access to the database itself, something that is often not applicable for hidden web databases. Even if we assume that we somehow have a full access to the database, finding exceptions of aggregates constantly requires formidable resources. Previous studies  \cite{online aggregation, Wavelets} proposed using precomputed samples of the data (e.g., uniform random sampling) to reduce response time while giving acceptable answers. Yet, selecting inappropriate samples for arbitrary aggregates would lead to large estimation errors. Therefore, selecting the right samples to give the right estimations and find exceptions while minimizing the query cost becomes a very complicated but important problem.\par

\section{Problem Definition}
The problem we consider in this paper is how to track and discover exceptions of aggregates over dynamic hidden web databases.\par

\subsection{Dynamic Hidden Databases} \label{hidden database}
In this paper, we restrict our discussion to categorical data. Consider a database $D$ with $n$ tuples $t_1, ..., t_n$, and $d$ attributes $A_1$, ..., $A_d$, each of which has a discrete domain. The domain of $A_i$ is denoted by $dom(A_i)$ for each $i$ $\in$ $[1: d]$, where $|dom(A_i)|$ represents the cardinality of $dom(A_i)$, i.e., the number of possible values of $A_i$ (domain values are always known). We assume no duplicate tuple exists in $D$, and each tuple $t_i$ can be represented by a d-dimension vector $(a_{i1} , a_{i2} , ..., a_{id} )$.\par
We assume a prototypical interface where users can query the database by specifying values for a subset of attributes. Thus a search query $q$ is of the form:\par
\noindent SELECT $*$ FROM $D$ WHERE $A_{i_1} = v_{i_1} \& ... \& A_{i_m} = v_{i_m}$, where $v_{i_j}$ is a value from $dom(A_{i_j})$, and $Sel(q)$ is the result of $q$.\par
Let $Sel(q)$ be the set of tuples in $D$ that satisfies $q$. As it is common with most web interfaces, we shall assume that the query interface is restricted to only return $k$ tuples, where $k \ll n$ is a predetermined small constant (such as 100 or 500). Thus, $Sel(q)$ will be entirely returned iff the number of tuples returned is less than $k$. If the query is too broad (i.e., number of tuples returned is more than $k$), only the $top-k$ tuples in $Sel(q)$ will be selected according to a ranking function, and returned as the query result. Note that repeating the same query $Sel(q)$ may not retrieve new tuples, i.e., the same $k$ tuples may always be returned. Along with $top-k$ tuples, we also assume that the interface returns the total number of tuples satisfying $Sel(q)$. Let $|q|$ be the total number of tuples returned by the interface, such that $|q|$ can determine the status of $q$; We say $Sel(q)$ is an overflow when $|q| > k$, i.e., that not all tuples satisfying $q$ can be returned. At the other extreme, when $q$ is too specific such that $|q| = 0$, we have an underflow, i.e., no tuples returned. If $0<|q|\leq k$ then we have a valid query.\par
For the purpose of this paper, we assume that a restrictive interface does not allow users to scroll through the complete answer of $Sel(q)$ when $q$ overflows. Instead, users must pose new queries by reformulating some of the search conditions. This is a reasonable assumption since many real-world $top-k$ interfaces (e.g., Google) limit  page turns (e.g., 100 at the time this paper was written).\par

\subsection{Aggregate Query}
Aggregate query $q$ for a target attribute $A_t$ with the selection condition $C$ takes the form of:\par
\noindent SELECT $Aggr(A_t)$ FROM $D$ WHERE $C$ \par
Where $Aggr$ is an aggregate, such as $COUNT$, $AVG$, $SUM$, $MIN$, and $MAX$. The selection condition $C$ is a conjunction of $A_{ij} = v_{ij}$ , where $v_{ij}$ is a value from $dom(A_{ij})$.\par
For example, assume we have two attributes $A_1$ and $A_2$, both are boolean attributes. $C$ could be a conjunction of $A_1 = 1$ AND $A_2 = 0$, and we donate the result of this aggregate query as $Aggr(q)$. Now let us say you want to know the average price of the HTC cellphones with a five-inch screen. The aggregate query takes the form of:\par
\noindent SELECT $AVG(price)$ FROM $cellphones$ WHERE $brand = ``HTC"$ AND $screenSize =  ``5"$

\subsection{Exception}\label{exception}
Consider an aggregate query, based on its values with fixed time interval at $T_0, T_1, ..., T_c$. We define an aggregate query as an exception by three factors. Firstly, the aggregate which is always over only a few of tuples is unnecessary to consider. For example, if only one tuple satisfies the selection condition, the change of itself will be the change of the aggregate. Therefore, this aggregate won't be as meaningful as others with a lot of tuples, so we set a $Support$ $Threshold$ $s$ to filter the small size aggregates. Thus, if $|q| < s$, we do not treat its aggregate as a potential exception. Secondly, if the aggregate is far away from what it was before, we consider it as a potential exception. To measure how ``far away" this aggregate is from what it was before, we compute the distribution of $Aggr(q)$ based on historical data. Then we can have an interval $(LB,UB)$, the average of $Aggr(q)$ $\mu \in (LB,UB)$, and the most of $Aggr(q)$ belong to $(LB,UB)$. If $Aggr(q)_{T_c}$ is out of $(LB,UB)$, it is a low probability event. Thus, we continue to treat it as a potential exception. To define the low probability event, we set a $Change$ $Threshold$ $p$, such that a low probability event has a chance of occurrence of less than or equal $p$ (or more than or equal $1-p$). Thirdly, the aggregate change percentage (increment/decrement) of last day comparing with previous days must be big enough. Thus, we set a $Percentage$ $Threshold$ $z$, such that an aggregate with a change percentage for last day is an exception when the change percentage is more than or equal $z$. Once an aggregate meets all these three factors, we treat it as an exception.\par
We define an aggregate as an exception at $T_c$ when it satisfies all conditions below:\par
(1) $max_{0 \leq i \leq c} (|q|_{T_i}) \geq s$;\par 
(2) $F_X(Aggr(q)_{T_c}) = P(X \leq Aggr(q)_{T_c}) > 1 - p$ or $F_X(Aggr(q)_{T_c}) = P(X \leq Aggr(q)_{T_c}) < p$;\par
(3) $\Delta(Aggr(q)_{T_c}) \geq z$;\par
Where $\Delta$ is the change percentage, $F_X(x)$ is the cumulative distribution function of $X$, and $X$ is a random variable such that $X \sim N(\mu, \sigma)$ in which $\mu$ is the average of $Aggr(q)_{T_i}$ and $\sigma$ is their variance $(i = 1, 2, ..., c - 1)$. Here, we assume that $Aggr(q)$ is in normal distribution form.\par
An example as follow: assume we want to monitor the average price of all cellphones, where $|Aggr(q)| > s$, $p = 0.01$, and $z = 15\%$. If the average price of the Motorola cellphones with 4-inch screen drops suddenly, based on the historical data, the current AVG($Price_{current}$) drops $15\%$ or more and in the left side of the normal distribution graph. $P(Price_{current}) < 0.01$. Thus, ``SELECT $AVG(Price)$ WHERE $brand$ = Motorola \& $ScreenSize$ = 4-inch" is returned as an exception.\par

\section{Solution}
A straightforward approach is to crawl the entire database $D$, then find exceptions. The benefit of this solution is that $all$ exceptions can be found. However, this is unreasonable solution for this type of problem since you don't only need to crawl $D$ in $T_c$ but also for all time interval $T_0, ..., T_{c - 1}$. Moreover, because of query limitation forced on hidden web databases, we cannot enumerate all the selection conditions to check whether each of them is an exception or not. A more reasonable approach is to take a sample to estimate the $Aggr(q)$. There is a deep research for answering this question for estimation with selection condition. But in this problem, estimating each selection condition separately wastes a lot of queries, because a sample can not only be for one selection condition, i.e., if we already know $Sel(q)$, where $A_1 = 0$  $\&$ $A_2 = 0$, $Sel(q)$ is a sample for both $A_1  = 0$ and $A_2 = 0$. For example, if we have the full list of the Motorola cellphones with 4-inch screen already, these cellphones can be a sample to estimate average price for both Motorola cellphones and 4-inch screen cellphones. Hence, we take a sample of the dataset first and then make different estimations based on it.\par
Now we introduce two concepts that are essential in our solution, Query-Pool and Apriori.\par
\noindent\textbf{Query-Pool:} A Query-Pool $QP$ is introduced to guide the process of identifying $candidate$ queries. A query $q$ is a candidate when its result $|q| \geq s$, i.e., the number of returned tuples is bigger than support threshold. Our purposed algorithms use $QP$ as part of their solutions to identify candidates. Once all candidates are identified and in $QP$, we answer aggregate queries and find exceptions.\par
\noindent\textbf{Apriori Algorithm:}\label{apriori} Apriori algorithm \cite{apriori} is a famous and influential algorithm for mining frequent itemsets for boolean association rules. One of its functionality is to determine the frequent itemsets, i.e., the sets of item which has minimum support. Apriori extends frequent subsets by one item at a time in a ``bottom up" approach (a step known as candidate generation) and tests the groups of candidates against the data. Similarly, the query candidates generation in our query-pool uses the same approach, where predicates are extended one attribute at a time until no further successful extensions are found. Recall that, as mentioned in \textbf{Query-Pool}, a query $q$ is a candidate when $|q| \geq s$. For example, assume we have two attributes $A_1$ and $A_2$, both are boolean, and $s = 100$. We have eight queries $q_1: (A_1 = 0)$, $q_2: (A_1 = 1)$, $q_3: (A_2 = 0)$, ..., $q_8: (A_1 = 1 \& A_2 = 1)$. A query $q_i$ for $i \in [1, 8]$ is a candidate when $|q_i| \geq 100$.\par

\subsection{Baseline Algorithm} \label{baseline}
The Baseline algorithm consists of two phases. The first one is generating the query-pool $QP$ using Apriori algorithm discussed in Section \ref{apriori}. Once this phase is done, we have all candidates in $QP$. The second phase is to find all exceptions from the query-pool. This can be done using the random walk approach to sample hidden web databases, which is proposed in \cite{random walk}. The idea of random walk is centered on $drill$ $downs$ over a query tree. The root level of query tree is SELECT $*$ FROM $D$, where the query tree organizes queries from broad to specific from top to bottom. For each $drill$ $down$, the query appends a random conjunctive constraint to the selection condition until a valid query is reached.\par
To imagine the random walk process, assume a specific ordering of all binary attributes each time, e.g. $[A_1, A_2, ..., A_d]$. The random walk starts from the root by issuing the query $q$ = SELECT * FROM $D$. Each time $q$ overflows, we expand it by adding the next attribute and assign a random predicate to it one at a time. For example, when $q$ overflows for the first time, we expand it by assigning a random selected predicate to $A_1$ (either ``$A_1 = 0$" or ``$A_1 = 1$"). This process leads $q$ to be either a valid or underflowing query. If $q$ is valid, then we randomly choose the returned tuple. Otherwise, we restart the random walk process.\par

Consider Figure 1, which shows a database $D$ with three attributes and three tuples, and a complete binary tree with 4 $(= d+1)$ levels. The $i$-$th$ level $(i \leq d)$ represents attribute $A_i$ and the leaves represent possible tuples. Each node that falls in a level $\leq d$ has two edges labelled as 0 and 1 respectively. The leaves at level $d$ + 1 represent all possible assignments of values to the attributes. The combination of any assignment is unique such that there is no other leave node with the same path. Not all leaves necessary represent existing tuples as some leaves may be empty, which is very common in a real-world database. When applying the random walk, we start from the top node of the tree $A_1$ (i.e., the first attribute) and randomly we pick a value either 0 or 1 (i.e., represents the edge) with equal probability. Every time we pick an attribute we check whether it is an overflow or not. If it is an overflow, we pick a random value for the next attribute, in this case it is $A_2$, and check again for its validity. We repeat this and assign attributes in order until we reach a valid or underflow query. If the query is underflow we start the process from the beginning, otherwise we have a valid query returning $k' \leq k$ tuples. We then we pick one of the $k'$ tuples with probability of $1/k'$.Note that the access probability of the tuple that gets picked is therefore $s(t)$ = $1 \over k' 2^{d(t)-1}$.  Then, the tuple is accepted with probability $a(t)$ where $a(t) = min\{Ck'2^{d(t)-1}, 1\}$ where C is a scale factor that boosts the selection probabilities to make the algorithm efficient. For categorical dataset, the only difference is that we have $|dom(A_i)|$ choices at $i$-$th$ level. And the access probability is $s(t) = 1 \over k' \prod_{i = d(t)}^{1} |dom(A_i)|$.\par
For each candidate in $QP$, we issue queries based on the approach above to get samples at $T_0, T_1, ..., T_c$. But instead of starting from SELECT $*$ FROM $D$ as the root, the root of each query candidate is the query candidate itself, i.e., if $q_i \in QP$ is SELECT $*$ FROM $D$ WHERE $A_1 = 0$ \& $A_2 = 0$, the root for this query tree is $A_1 = 0$ \& $A_2 = 0$. Based on our definition in Section \ref{hidden database}, the candidate starts with extremely broad (and thus overflowing) query and the drill-down process narrows the query down by adding randomly selected predicates, until a valid query is reached. Once we have a valid query, we select our sample. \cite{random walk} describes the random walk process in more details.\par
When selecting a sample, we also compute the probability $p(q)$ of this sample being selected in a drill-down. Same sample from different candidates could have different $p(q)$ based on the number of drill-down levels performed before selecting the sample. To illustrate this, consider two candidates $c_1$ and $c_2$, such that $c_1$ is $A_1 = 0 \& A_2 = 1$ and $c_2$ is $A_1 = 0 \& A_2 = 1 \& A_3 = 0$. If both $c_1$ and $c_2$ select the same sample $s_1 (A_1 = 0 \& A_2 = 1 \& A_3 = 0 \& A_4 = 1 \& A_5 = 0)$, their probabilities will be different, $1 \over 8$ and $1 \over 4$ respectively (i.e., $c_1$ has three drill-down levels while $c_2$ has two drill-down levels). Once we select the sample and calculate its probability, we estimate the average price for the candidate and judge whether $Aggr(q)$ is an exception or not by the definition in Section \ref{exception}. Thus, the algorithm is as follow:\par

\begin{algorithm}[htb]
\caption{Baseline}
\label{alg:baseline}
 \KwData{Workload $W$}
 \KwResult{A sample set and the estimations for $W$}
 \For{$i=0; i<n; i++$}{
    $q \gets c_i$ node.
    
    $S_i \gets$ take new samples for $c_i$;
    
    \If{$S_i$ is exception}
    {
        return $c_i$ as exception;
    }
}
 
 $n$ is total number of candidates in $W$.
 
 $c$ is a candidate.
 
 $S$ is set of samples corresponding to $c$.
 
\end{algorithm}

\vspace{1mm}

However, a sample which satisfies the selected conditions of several candidates in $QP$ at the same time can only be used to estimate one of the candidates once by this algorithm. For example, the sample $s_1$ above can be used for both candidates $c_1$ and $c_2$, but by the baseline algorithm, it is only used once. When the intersection of two candidates is very big, this is obviously not a good idea. To estimate the aggregates of all candidates over time, the baseline algorithm treats every candidate on $T_0$, $T_1$, $...$, $T_c$ separately. The estimations are independent with each others. Although it is simple, there are two obvious disadvantages. Firstly, it wastes numerous queries because no information is retained/reused from time to time. Secondly, when there are two candidates having an intersection with each other, it is a waste that the samples in the intersection are only for one estimation as the naive algorithm. In subsections \ref{stratified-detector} and \ref{udometer}, we introduce our STRATIFIED-DETECTOR and UDOMETER algorithms to address these two problems based on the state-of-art techniques separately.\par

\subsection{Our Algorithms} \label{our algorithms}
\subsubsection{STRATIFIED-DETECTOR}\label{stratified-detector}
Given a query-pool $QP$, STRATIFIED-DETECTOR improves the selectivity of samples for this query-pool. The $Stratified$ $Sampling$ technique reduces query cost while minimizing estimation error for a given query-pool, which is proposed in \cite{stratified sampling}. The idea of stratified sampling is to generalize uniform sampling by partitioning the population into multiple strata and samples are selected uniformly from each stratum. The more ?important? the strata is, the more contribution it has on picking samples.\par
\noindent\textbf{Stratified Sampling}\newline
\noindent\textbf{Example:} Let us consider a database $D$ that has one attribute $price$ and four tuples, such that $t_1$ = \{100\}, $t_2$ = \{150\}, $t_3$ = \{200\}, and $t_1$ = \{250\}. Let us have the aggregate query $q_1$ = SELECT COUNT($*$) FROM $D$ WHERE $price \geq 200$. Let P({$q_1$}) defines the population of $q_1$ on $D$ as a set of size $|D|$ that contains the value of the aggregated column selected by $q_1$, or 0 otherwise. We have P({$q_1$}) = \{0, 0, 1, 1\}. Since P({$q_1$}) has a mix of 1's and 0's, it is a nonzero variance, which makes uniform sampling a poor choice for this problem. To have a zero variance, we better partition $D$ into two starta \{t1, t2\} and \{t3, t4\}, such that P({$q_1$}) contains now two starta \{0, 0\} and \{1, 1\}, with both have zero variance. Nonetheless, if we also have $q_2$ =  SELECT COUNT($*$) FROM $D$ WHERE $price \geq 150$, then P{$q_2$} = \{0, 1, 1, 1\} (different than P({$q_1$})),  and each query will have its own population of $D$. So the challenge is to adapt stratified sampling so that it works well for all queries in query-pool.\par
In general, stratified sampling partitions $D$ into $r$ strata with $t_1, ..., t_r$ tuples (where $\sum t_j = n$) with $m_1, ..., m_r$ tuples uniformly sampled from each stratum (where $\sum m_j = m$). The stratified sampling proposed in \cite{stratified sampling} discusses how can we be apply stratified sampling effectively in databases. It consists of three steps (1) \textit{stratification}, to determine the number of strata $r$ to partition $D$ into and the number tuples from $D$ for each stratum, (2) \textit{allocation}, to determine how to divide $m$ into $m_1, ... m_r$ across $r$ strata, (3) \textit{sampling}, apply random walk to sample $m_j$ tuples from stratum $D_j$.\par
When applying the steps above into our problem, we divide the query-pool into starta, such that the number of these starta $F = \{D_1, ..., D_r\}$ where for any starta $D_j \in F$, each $q_i \in QP$ selects either $all$ tuples in $D_j$ or none. The number of starta $r$ depends on both $D$ and $QP$ (generally, the total number is upper-bounded by $min(2^{|W|}, n)$). After identifying starta, stratified sampling algorithm picks samples. Note that the $r$ strata must be non-overlapping and mutually exclusive.\par
However, the above technique is not suitable when the size of query-pool is big (i.e., number of starta is large). Instead, when the space of one candidate $c_2$ is the subset of another $c_1$, STRATIFIED-DETECTOR takes samples separately from $c_2$ and $c_1-c_2$, then merge these two together by stratified sampling to estimate $c_1$. Here we can get the estimations for both $c_2$ and $c_1$. Note by applying this technique, we also need to use the probability of $c_2$ to calculate $c_1$. Further more, if we have another candidate $c_3$, which $c_1$ is a subset of $c_3$, we can directly take sample from $c_3-c_1$, and then get the estimation of $c_3$ by the sample of $c_3-c_1$, and the stratified sample of $c_1$ we got previously. Moreover, when a candidate $c_1$ contains more than one longer candidate, we consider the longer candidate with the biggest subspace in term of size (i.e., the longer candidate with biggest COUNT). For example, if $c_1$ contains three longer candidates $c_{1_1}$, $c_{1_2}$, and $c_{1_3}$ with COUNT of 1000, 2000, and 3000 respectively, we use $c_{1_3}$ to estimate $c_1$.\par
To illustrate this stratified sampling, let us reconsider $c_1$ and $c_2$, such that $c_1$ is a two conjunction candidate and $c_2$ is a three conjunction candidate. Note that $c_2$ exists in $c_1$. Thus, the average price of $c_1$ is\par
\noindent $AVG(c_1) = (x*COUNT(c_1-c_2)+y*COUNT(c_2))/COUNT(c_1)$\par
Where $x$ is the average price of $(c_1-c_2)$ and $y$ is the average price of $c_2$. If we select the samples $u$ (level 2) and $v$ (level 3) to calculate $y$, then\\par
\noindent $y = {{u \over p(u)} + {v \over p(v)} \over {1 \over p(u)} + {1 \over (p(v)}}$\par
\noindent Note that $p(u) = |dom(A_3)| * |dom(A_4)|$ and $p(v) = |dom(A_3)| * |dom(A_4)| * |dom(A_5)|$.\par
Similarly, we can calculate $x$ where the only difference is that the $|dom(A_3)|$ becomes  $|dom(A_3) - 1|$ (i.e., $p(u)=|dom(A_3)-1|*dom|A_4| and p(v)=|dom(A_3)-1|*|dom(A_4)|*|dom(A_5)|$). Algorithm \ref{alg:stratified-detector} depicts the pseudocode.\par

\begin{algorithm}[htbp]
\caption{STRATIFIED-DETECTOR}
\label{alg:stratified-detector}
 \KwData{Query-Pool $QP$}
 \KwResult{A sample set and the estimations for $QP$}
 $L_{(k+1)}$~ is~ empty~ in $QP$\;
 \For{$i=k; k>0; k--$}{
    \For{$c \in L_i$~}{
        \If{$c$ does not include any $c\_one\_longer$ in $L_{(i+1)}$~}{
            take a new sample for $c$;
        }
        \Else{
            stratified sampling based on the count;
            
            take a new sample for $c$;
        }
    }
 }
 
 $L_k$ is the set of candidates with length $k$ workload.
 
 $c$ is a candidate.
\end{algorithm}
\noindent The details of the stratified sampling is:\par
For the current candidate $c_i$, take the $c\_one\_longer$ that contains the most tuples as $B$. Assuming that we have samples already for $c_j$, then take samples from the remaining subspace $c_i-c_j$ and then combing them with the sample in $c\_one\_longer$. And then we can have an estimation of the average of the target attribute, $AVG(c_i)=(AVG(c_i-c_j)*|c_i-c_j|+AVG(c_j)*|c_j|)/|c_i|$. $|c_i|$ is the count of tuples that satisfying the selected condition of the candidate $c_i$. After we finished the loop of $L_1$, we can have estimations for all the candidates in $QP$. After repeated on $T_0 , ..., T_n$, because we have the estimations for all the candidates on $T_i (i=0,1,...,n)$, we can find out which ones are exceptions based on the definition.\par
However, when the variance of the samples is big, we cannot have a good result. The reason is, when we randomly pick samples for a candidate on $T_0 , ..., T_n$,  with the big variance, we cannot distinguish whether a dramatic change is caused by the sampling or the data itself. A straightforward solution is to take more than one sample for each candidate in each time interval in order to reduce the variance (if the change is caused by sampling). The more samples we collect, the more accurate estimations we get. Of course, this solution is not practical neither reasonable since we have a limited number of queries to issue.\par
To overcome this problem, we introduce our UDOMETER algorithm discussed in the next subsection.
%

\subsubsection{UDOMETER}\label{udometer}
Even though most of the real-world databases are dynamic (updated at arbitrary time), yet many of these databases do not see frequent updates. And the fewer changes happen to the database, the few changes the sample will see. This means that the random walk technique discussed in Section \ref{baseline} could lead to a significant waste of queries after a period of time since no information is retained/reused. Not only that, but the saving of query cost can be directly translated to more accurate aggregate estimations. In particular, since updating a drill-down may consume fewer queries, the remaining query budget (after updating all previous drill downs) can be used to initiate new drill downs, increasing the number of drill downs and thereby reducing the estimation error.\par
To understand how saving query cost reduces estimation error, consider the example mentioned in Section \ref{baseline} (Figure \ref{fig:random walk}). Note that each drill down has a unique sequence number of the leaf-level node corresponding
to it, which can be uniquely identified this drill down. We notify this as a $signature$, such that at each time interval $T_i$ where $T_0 \leq T_i \leq T_c$, we have a signature set $S = \{r_1, . . . , r_h\}$  where each $r_i$ defines one drill-down performed. Now, to collect any sample (from Figure 1) at $T_0$, it requires at least 4 queries, i.e., from root to any leaf node (a tuple). If no change happens to the database at $T_1$, we can reach the same leaf node from $T_0$ in only 1 query (by issuing the valid query from $T_0$). This means we can save 3 out of 4 queries. However, we still need to issue one more query over leaf node's parent to determine if it is still the top non-overflowing query. This reduces the number of drill downs that can be updated at each interval time to at most half of the query budget. Moreover, the estimation produced in different time interval are no longer independent of each other due to the reuse of the same signature set (of drill-downs).\par
UDOMETER addresses the problem above by reissuing the same sample set from $T_0$ for all time interval $T_i$ where $T_0 \leq T_i \leq T_c$. Although, UDOMETER does not reduce the query cost for generating the query-pool (as Apriori still does it). Yet, it still minimizes the total query cost by reducing the query cost of the random walk process (along with reducing the estimation error). As for minimizing the total query cost even further for a given query-pool of aggregate queries, we still need to minimize the query cost of generating candidates.\par
When applying UDOMETER, we divide the problem into two parts based in the time interval: (1) Day one (i.e., $T_0$), (2) Day two and beyond $(i.e., T_1 , ..., T_n)$. For day one (i.e., $T_0$), we apply the STRATIFIED-DETECTOR as is (Algorithm \ref{alg:stratified-detector}). Once finished, we will have a sample set for $T_0$. Now for day two and beyond (i.e., $T_1, ..., T_n$), we reissue the same sample set from $T_0$. Note by doing so, each query-pool $QP$ in $T_1, ..., T_n$ is a subspace of the query-pool $QP_0$ (from day one $T_0$). Although the number of candidates that is considered by the UDOMETER may be less than the actual number of candidates for a specific day, the accuracy of our estimations should be higher, which lead to more accurate exceptions. Algorithm \ref{alg:udometer} provides the pseudocode for UDOMETER.\par

\begin{algorithm}[htb]
\caption{UDOMETER}
\label{alg:udometer}
 \KwData{Query-Pool $QP$}
 \KwResult{A sample set and the estimations for $QP$}
 $L_{(k+1)}$~ is~ empty~ in $QP$\;
 \For{$T_0$}{
    apply STRATIFIED-DETECTOR (Algorithm \ref{alg:stratified-detector});
    
    store samples as the sample of $c$;
}
\For{$T_1$ to $T_n$}{
    reissue same sample $c$ from $T_0$;
}
\end{algorithm}
Reissuing the same set of samples decreases the variance of the samples and lead to more accurate estimation. UDOMETER is now able to detect the overall trend of the average price of most candidates (whether its an increment or decrement). Yet, detecting the overall trend is sometimes not enough to detect the exception itself. Two of the main reasons that an exception occurs for a candidate are (1) sudden drop/rise in price for sizable number of tuples, and (2) insertion or deletion of new and existing tuples. When we consider a subspace $B$ with the most tuples to estimate $A$, these new or deleted tuples may not have a great impact on $A$ to be an exception because of the ratio size of $B$ to $A$. Our PRIORITY-UDOMETER, which we introduce in the next subsection can handle both events.\par

\subsubsection{PRIORITY-UDOMETER}
In the previous sections, we mentioned that when we applied the STRATIFIED-DETECTOR, we divided the target candidate $A$ into two subspaces $B$ and $A-B$, in which the count of $B$ is the biggest among the candidates which are the subspaces of $A$. It is a straightforward idea because we only need to take a new sample in the smallest space $A-B$. However, while this method to divide the space is good to estimate the aggregates like average and sum, it cannot achieve a good result for difference operator, like the change of average. For example, we have two candidates $B$ and $B'$, $B$ is the bigger subspace but the average is almost kept constant, while $B'$ is slightly smaller than $B$, but the average changes dramatically on $T_n$, which leads $A$ to be an exception. In this situation,  $B'$ and  $A-B'$ is obviously a better partition. On the contrary, if $B'$ is very small, although the average changes dramatically on $T_n$, it has little impact on $A$, $B'$ and  $A-B'$ cannot be a good partition. Thus, we purpose a score function $S(COUNT, Aggr(q)_{T_0}, ..., Aggr(q)_{T_n})$ to decide which one is the best partition.

The purpose of the score function is to compare the impact of different candidates which are the subspace of the target $A$. $A$ can be impacted by two factors, the size of the subspace $B$, and how much the aggregate changed. The score function on $T_n$ we propose is
\begin{equation}
\begin{split}
    S(COUNT, Aggr(q)_{T_0}, ..., Aggr(q)_{T_n})\\
    =abs(Aggr(q)_{T_n}-\sum_{i=0}^{n-1}{Aggr(q)_{T_i}})*\frac{COUNT(q)}{COUNT(A)}
\end{split}
\end{equation}
and we compute the score of each candidate which is a subspace of target $A$, then choose the one with the biggest score as $B$.\par

\section{Experiment}

\begin{figure*}[ht]
  \begin{minipage}[t]{0.24\textwidth}
    \includegraphics[width=\textwidth]{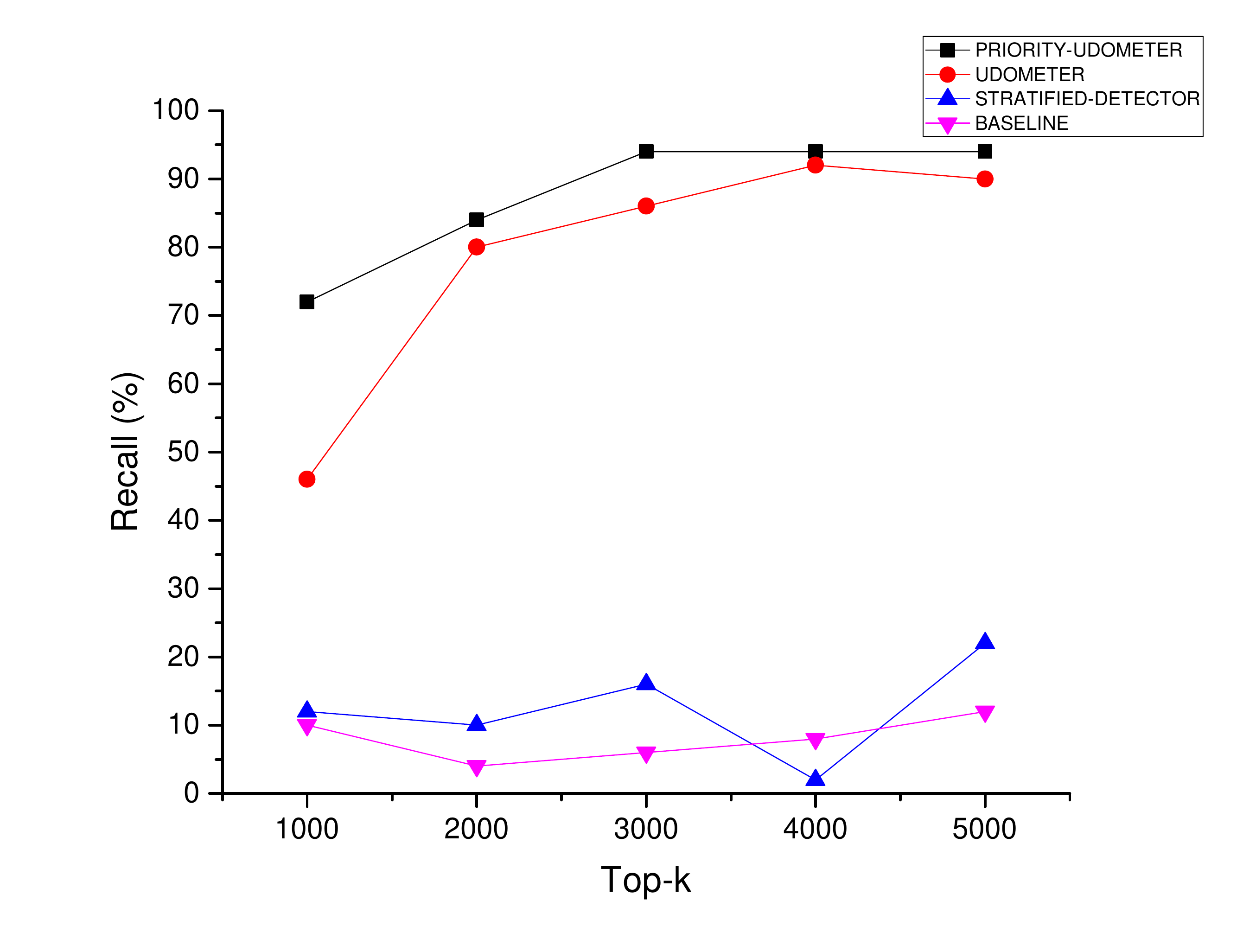}
    \caption{\small \textbf{Effect of $k$}}
    \label{fig:recallK}
  \end{minipage}
  \hspace{0.5mm}
  \begin{minipage}[t]{0.24\textwidth}
    \includegraphics[width=\textwidth]{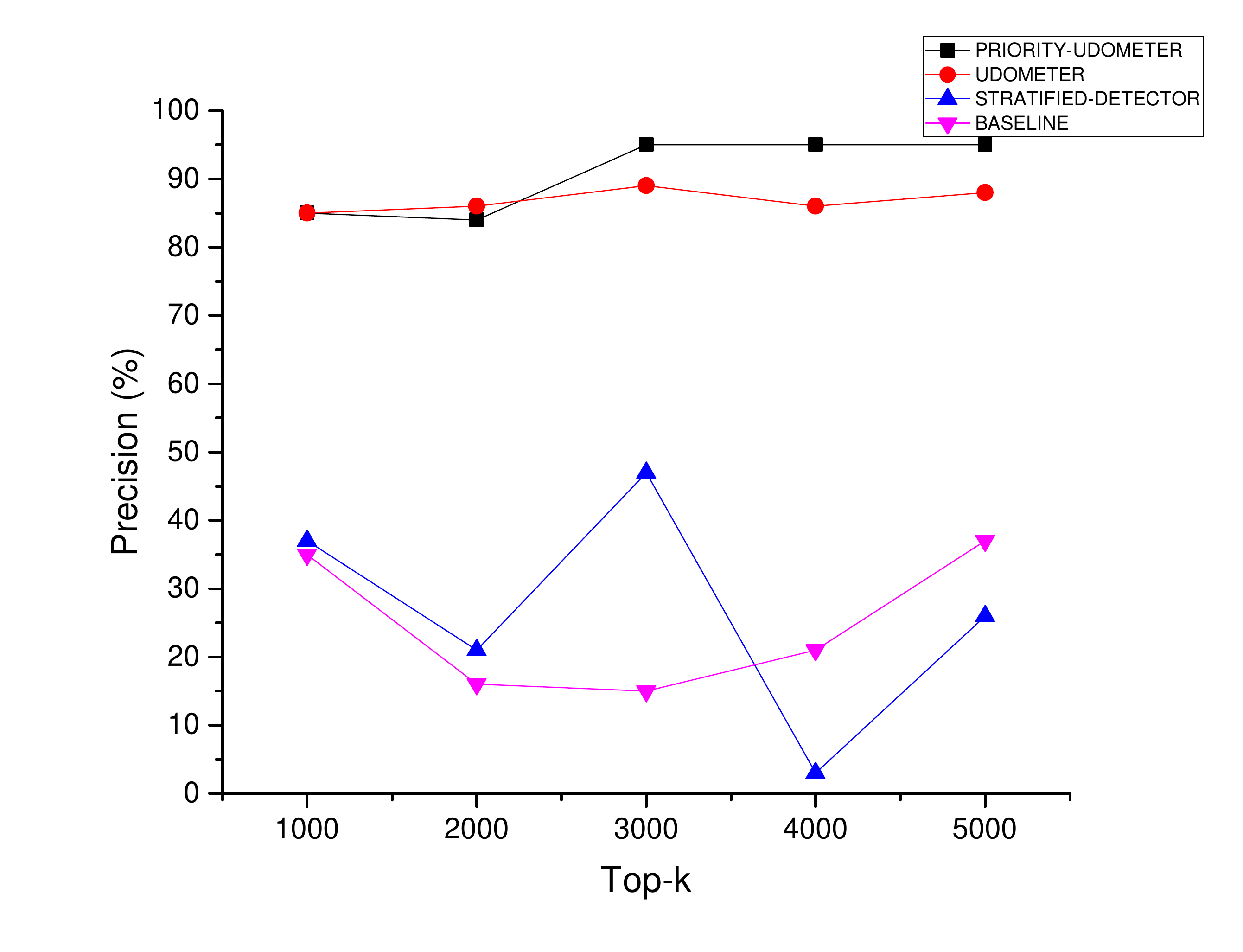}
    \caption{\small \textbf{Effect of $k$}}
    \label{fig:precisionK}
  \end{minipage}
  \hspace{0.5mm}
  \begin{minipage}[t]{0.24\textwidth}
    \includegraphics[width=\textwidth]{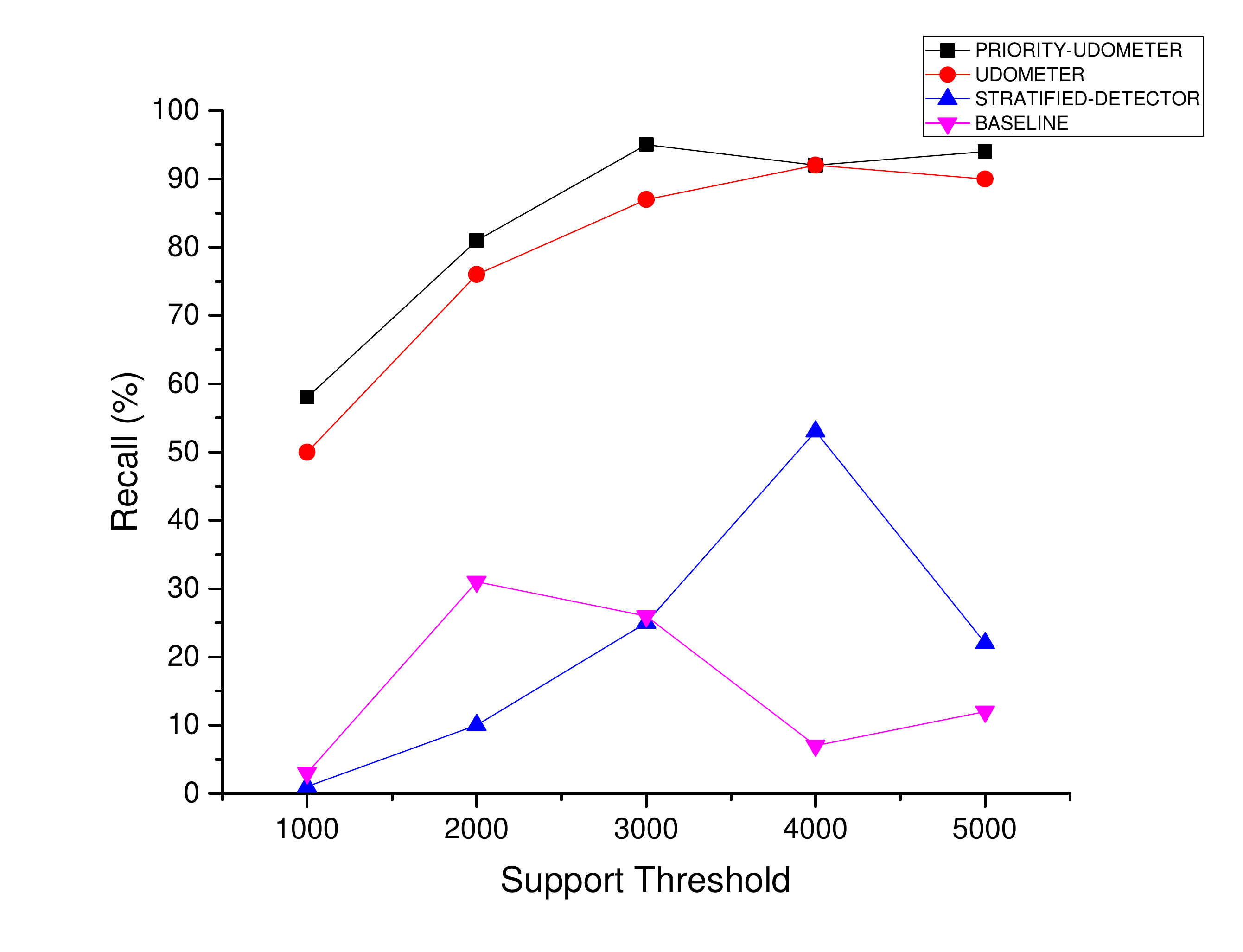}
    \caption{\small \textbf{Effect of $s$}}
    \label{fig:recallS}
  \end{minipage}
  \hspace{0.5mm}
  \begin{minipage}[t]{0.24\textwidth}
    \includegraphics[width=\textwidth]{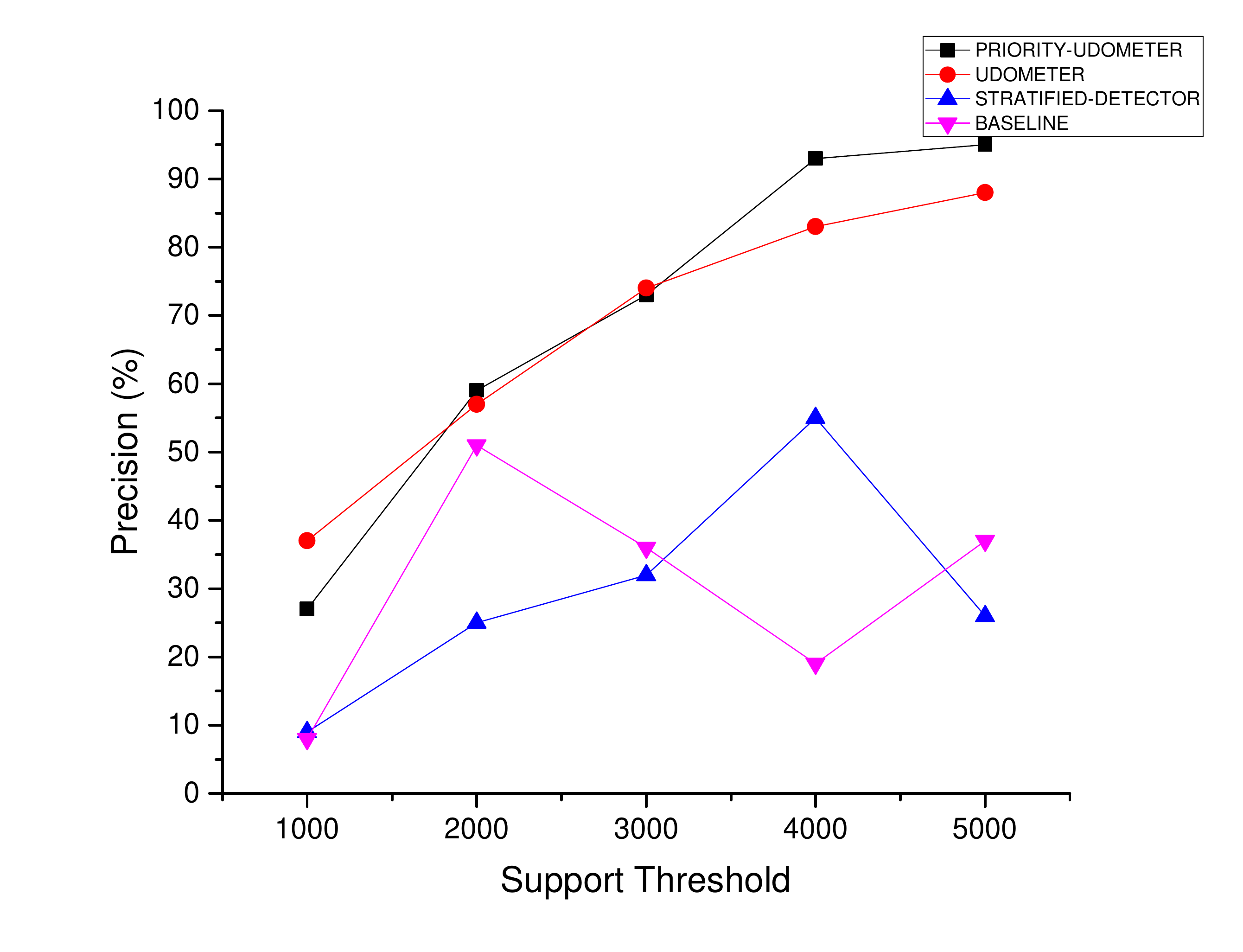}
    \caption{\small \textbf{Effect of $s$}}
    \label{fig:precisionS}
\end{minipage}
\end{figure*}

\begin{figure*}[ht]
  \begin{minipage}[t]{0.24\textwidth}
    \includegraphics[width=\textwidth]{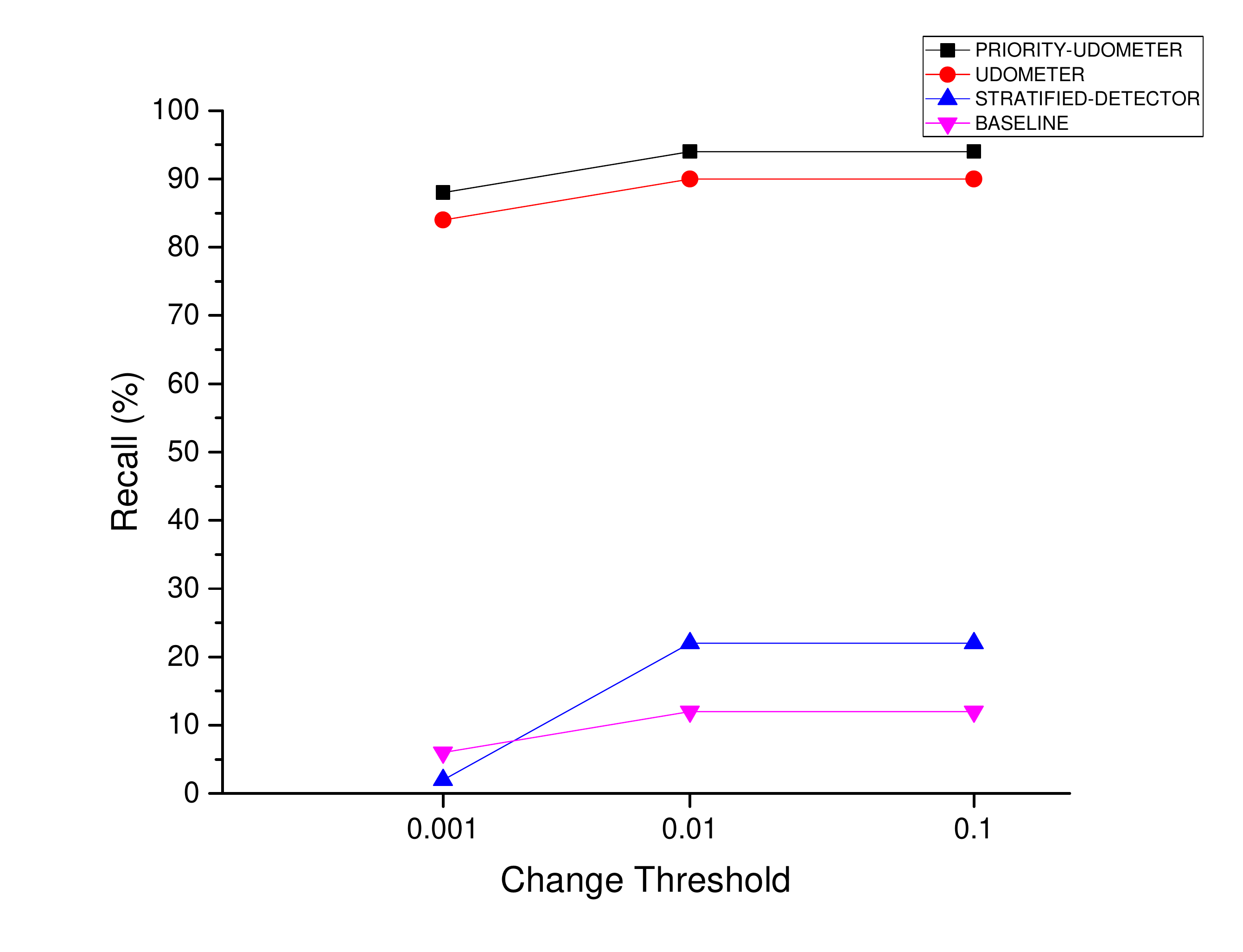}
    \caption{\small \textbf{Effect of $p$}}
    \label{fig:recallP}
  \end{minipage}
  \hspace{0.5mm}
  \begin{minipage}[t]{0.24\textwidth}
    \includegraphics[width=\textwidth]{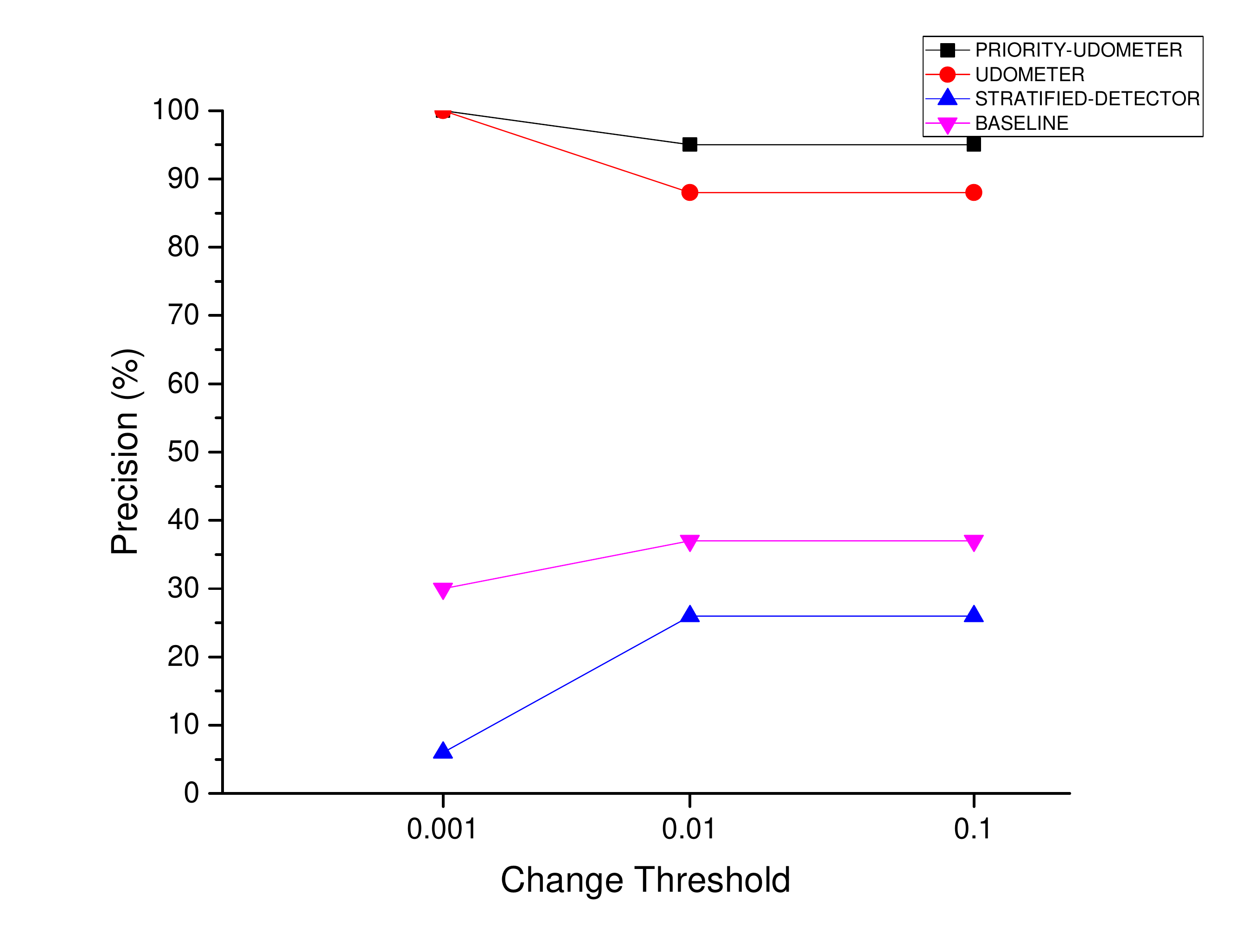}
    \caption{\small \textbf{Effect of $p$}}
    \label{fig:precisionP}
  \end{minipage}
  \hspace{0.5mm}
  \begin{minipage}[t]{0.24\textwidth}
    \includegraphics[width=\textwidth]{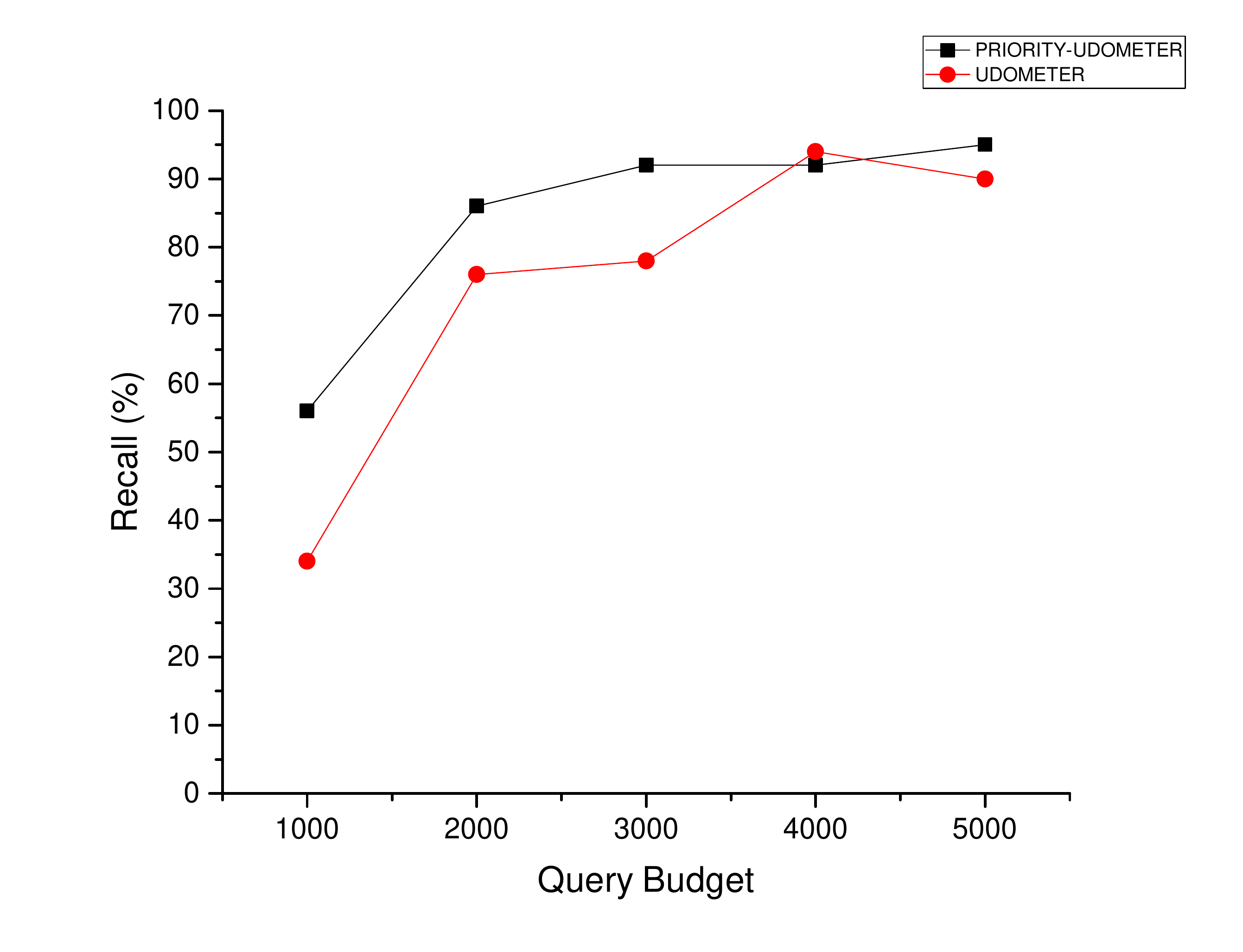}
    \caption{\small \textbf{Effect of query budget on recall.}}
    \label{fig:recallQB}
  \end{minipage}
  \hspace{0.5mm}
  \begin{minipage}[t]{0.24\textwidth}
    \includegraphics[width=\textwidth]{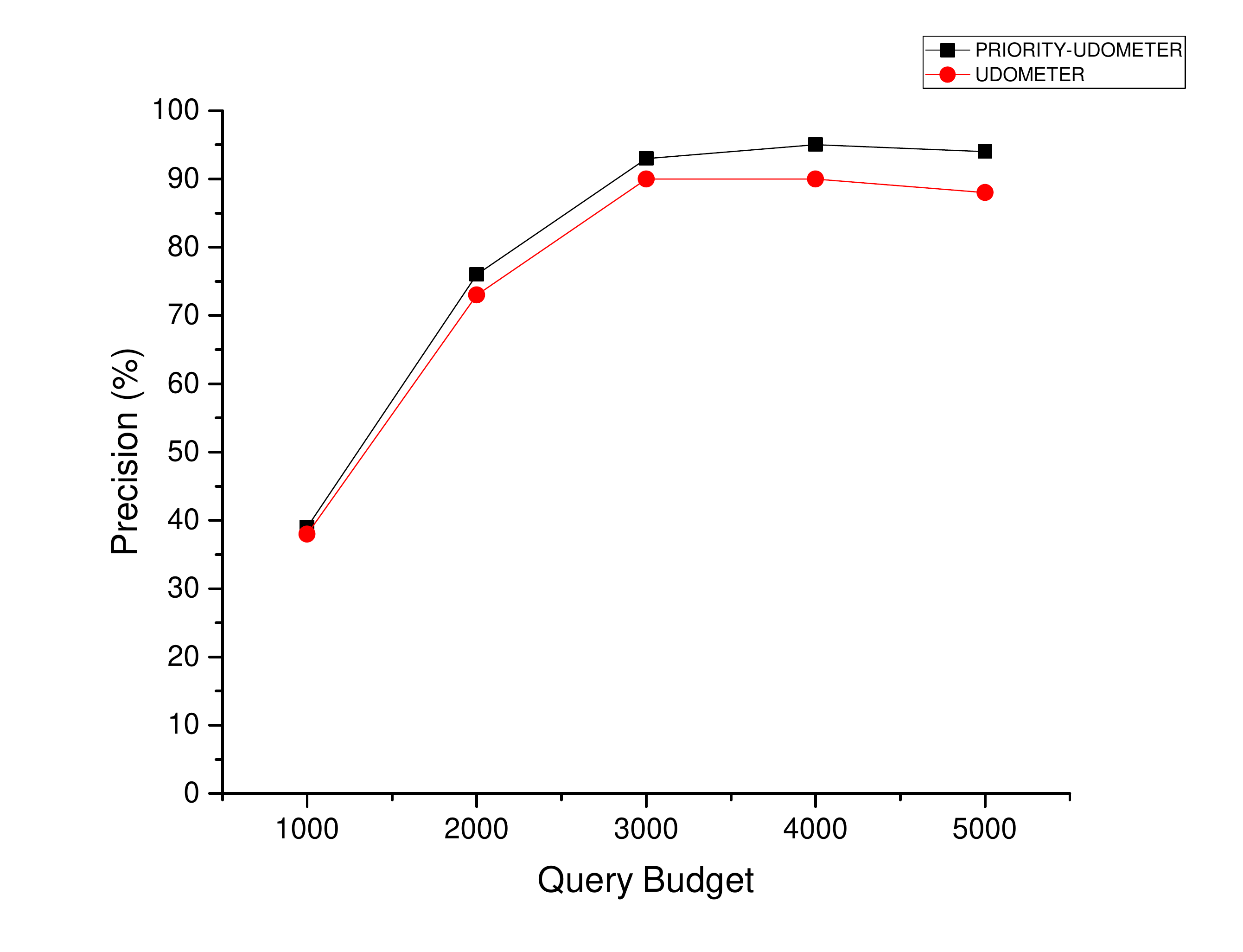}
    \caption{\small \textbf{Effect of query budget on precision.}}
    \label{fig:precisionQB}
  \end{minipage}
  \hspace{0.5mm}
  \begin{minipage}[t]{0.24\textwidth}
    \includegraphics[width=\textwidth]{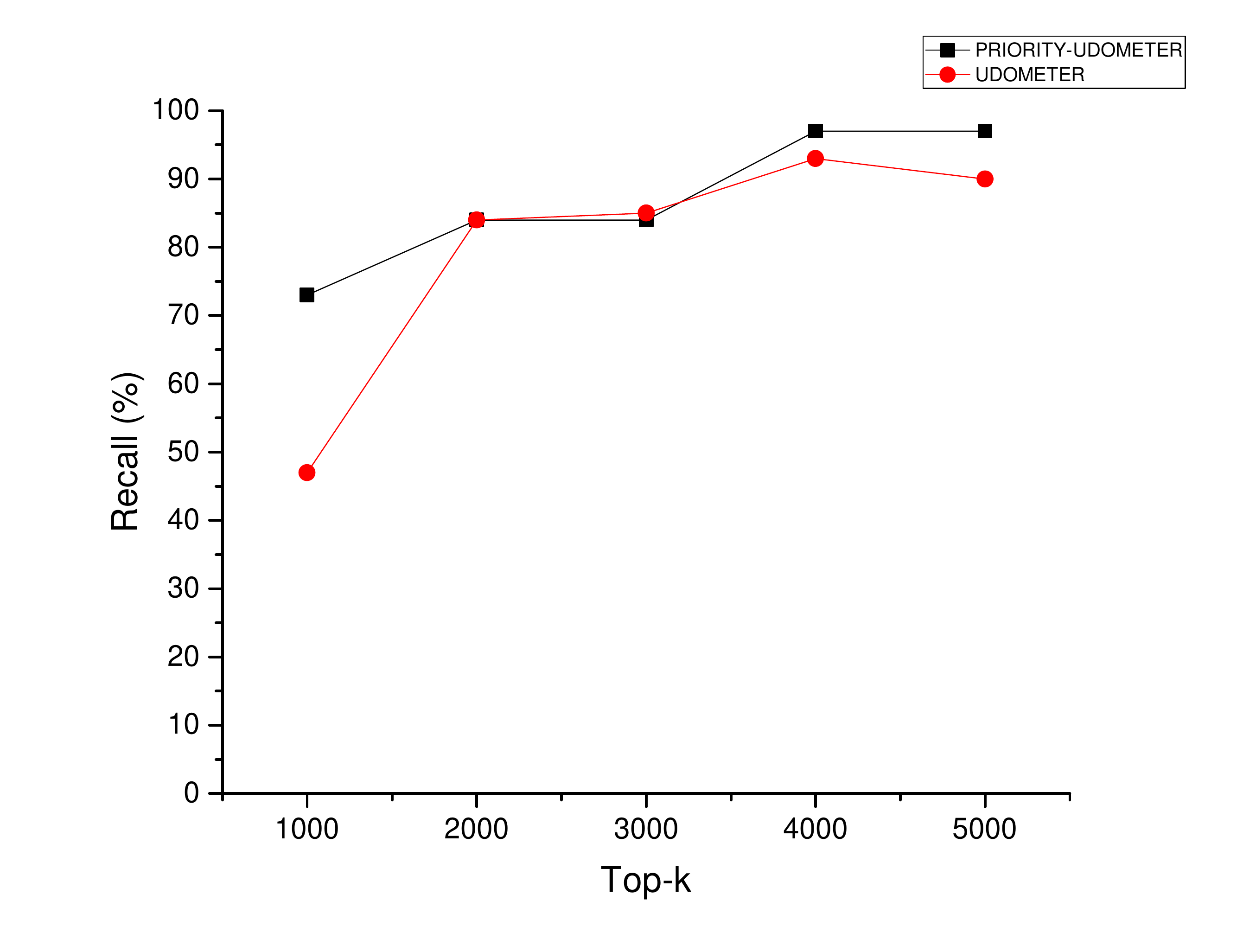}
    \caption{\small \textbf{Effect of $k$ with score function $S$.}}
    \label{fig:recallScoreFK}
  \end{minipage}
  \hspace{0.5mm}
  \begin{minipage}[t]{0.24\textwidth}
    \includegraphics[width=\textwidth]{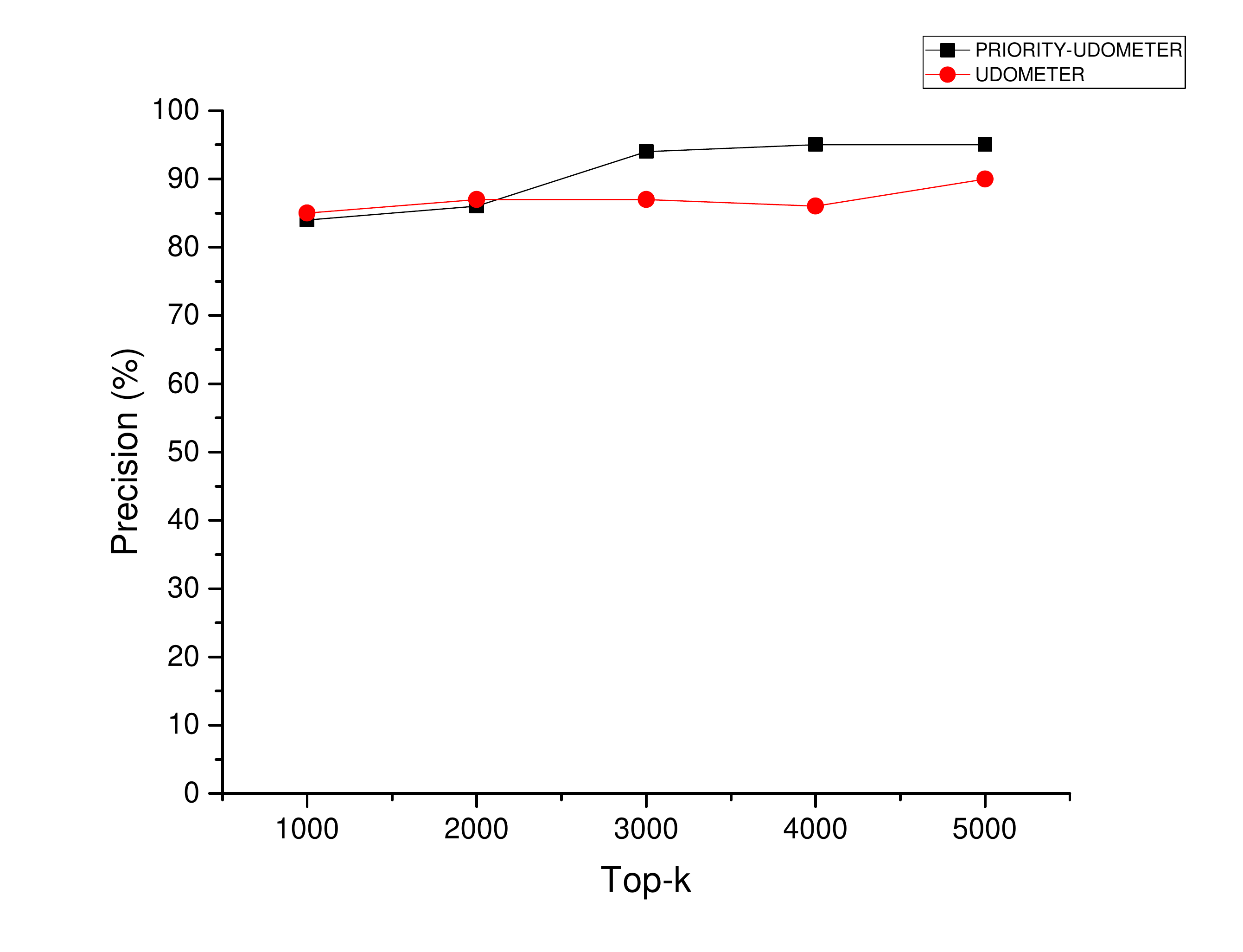}
    \caption{\small \textbf{Effect of $k$ with score function $S$.}}
    \label{fig:precisionScoreFK}
  \end{minipage}
  \hspace{0.5mm}
  \begin{minipage}[t]{0.24\textwidth}
    \includegraphics[width=\textwidth]{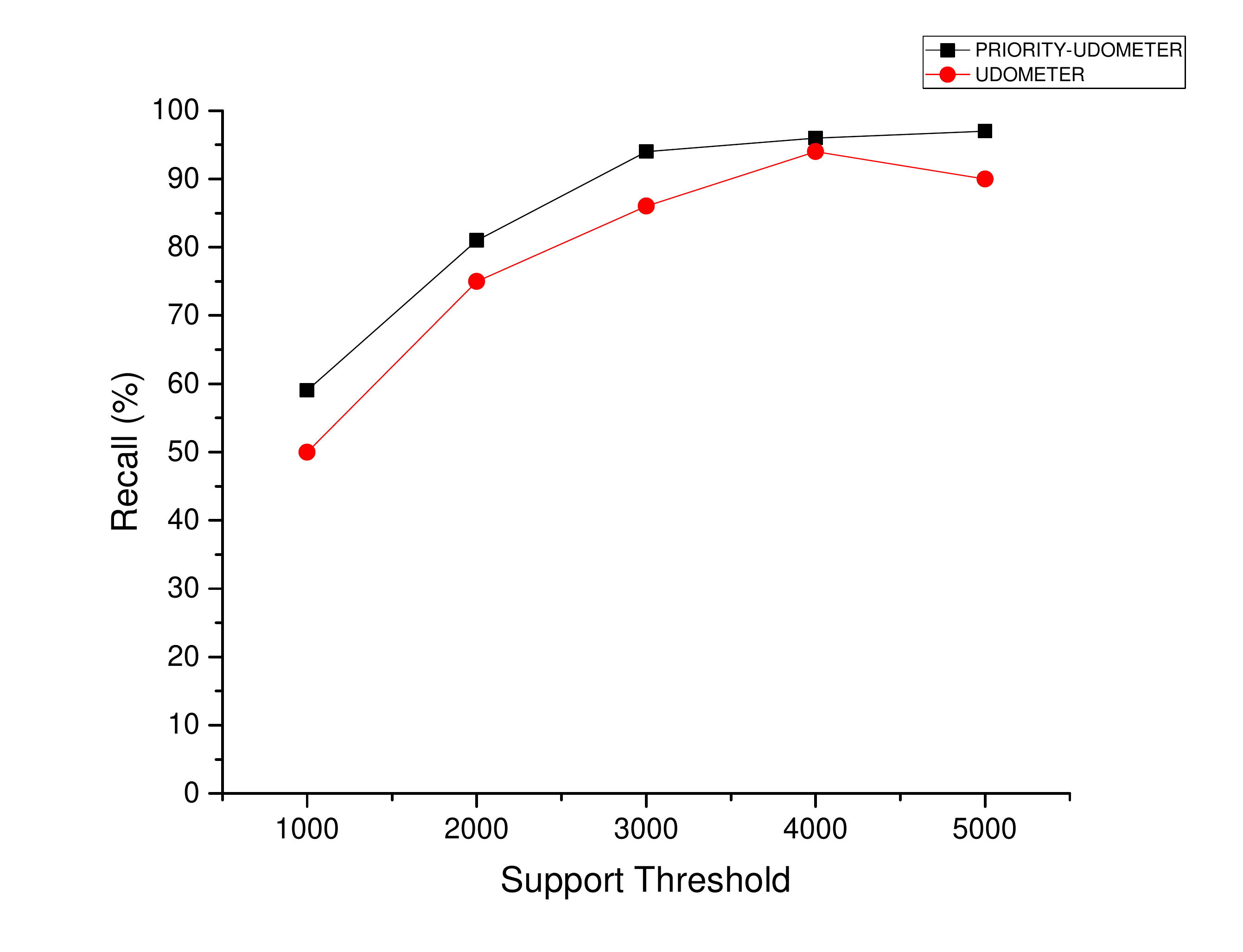}
    \caption{\small \textbf{Effect of $s$ with score function $S$.}}
    \label{fig:recallScoreFS}
  \end{minipage}
  \hspace{0.5mm}
  \begin{minipage}[t]{0.24\textwidth}
    \includegraphics[width=\textwidth]{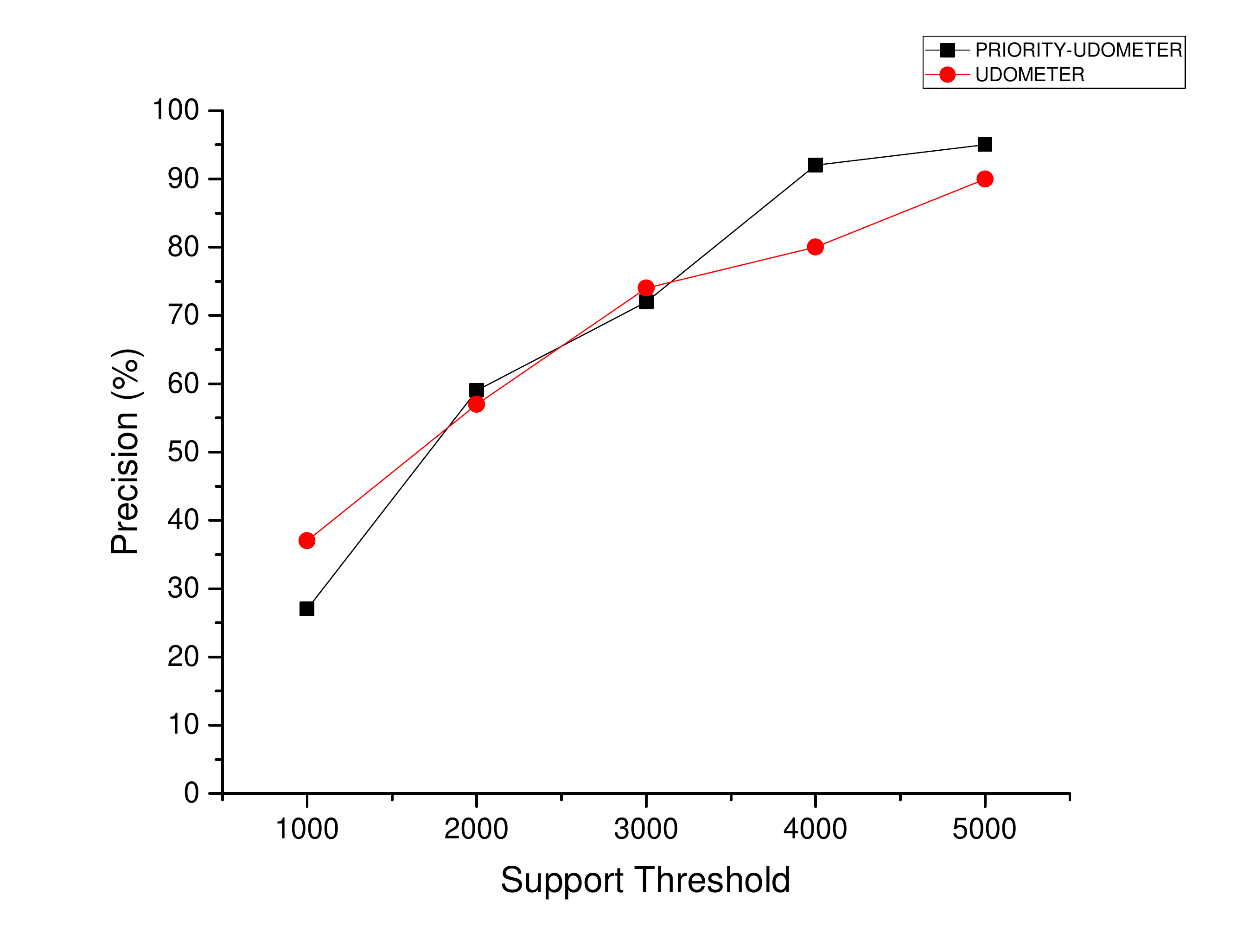}
    \caption{\small \textbf{Effect of $s$ with score function $S$.}}
    \label{fig:precisionScoreFS}
  \end{minipage}
  \hspace{0.5mm}
  \begin{minipage}[t]{0.24\textwidth}
    \includegraphics[width=\textwidth]{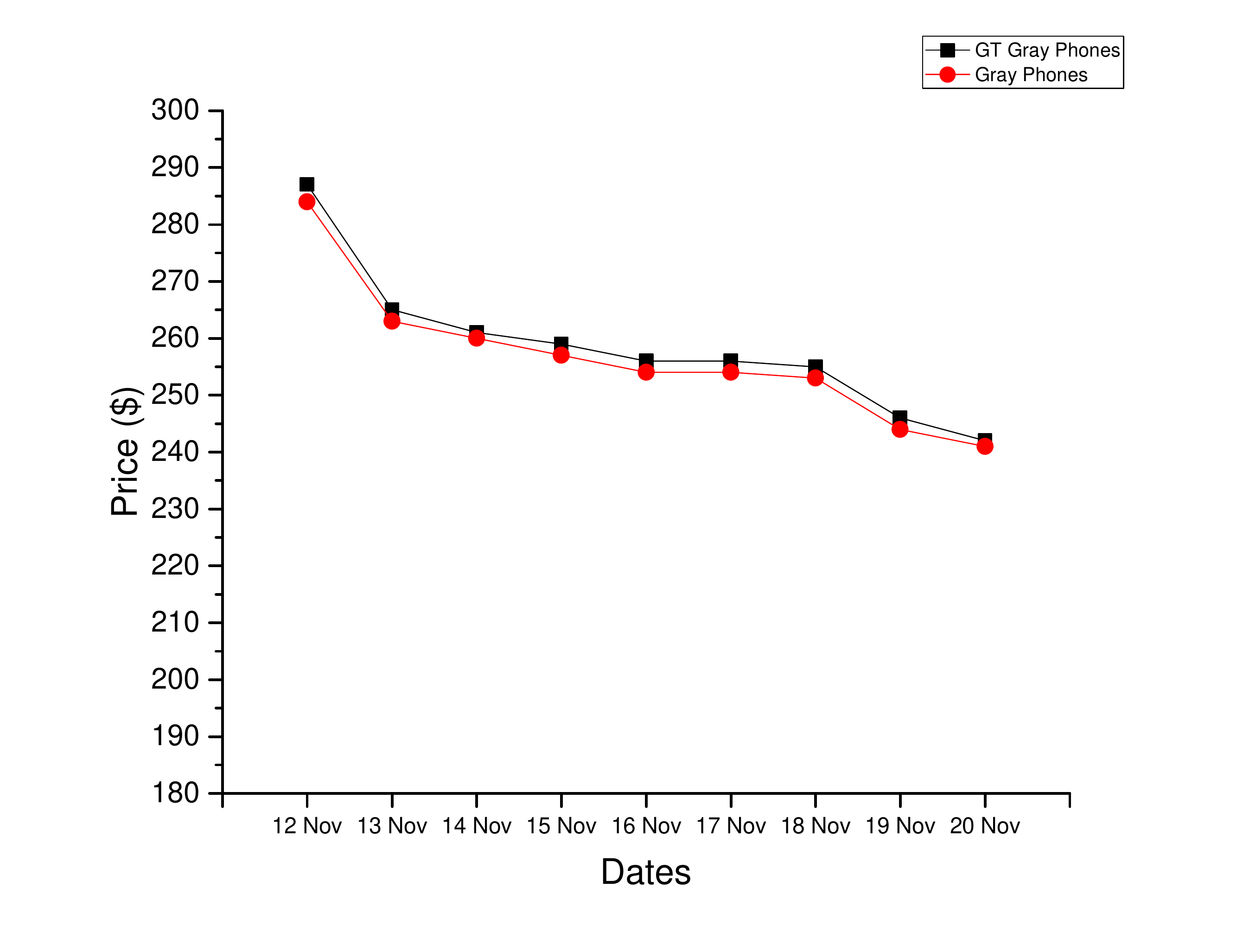}
    \caption{\small \textbf{Gray phones average price}}
    \label{fig:trend}
  \end{minipage}
\end{figure*}

\subsection{Experimental Setup}
\noindent\textbf{Dataset:} We tested our algorithm over a real world (categorical) web database eBay.com, specifically Cell Phones and Smartphones category, to which we have full offline access. The dataset, which we crawled, is for the period from Oct 21st, 2015 to Nov 20th, 2015 and contains 2,393,361 tuples and 8 attributes. All tuples that have been crawled offer a ``Fixed Price'' option while the attribute domain sizes ranges from 4 to 82.\par
\noindent\textbf{Challenges:} Recall from Section \ref{intro} that one of the main challenges hidden web databases is the query budget available daily. Even though we have a complete access to the dataset, we still applied same number of query budget provided by eBay.com per-IP and per-developer. The default query budget per day is 5000.\par
However, because of query limitation, our algorithms first use the query budget available for one day as a bootstrapping to generate the query-pool before starting the algorithms the following day.\par
\noindent\textbf{Query-Pool:} Examining all candidates in a query-pool is very challenging because of the limitation of query budget. Bigger query-pool means less query budget we can spend per candidate. We tested our four algorithms under 5 different sizes of query-pool (i.e., with supported threshold $s$ $\geq$ 1000, 2000, 3000, 4000, and 5000). Based on our dataset, this means that we sometimes can only issue two query budget per-candidate, while other times we have the luxury of spending sixteen query budget per-candidate.\par
\noindent\textbf{Releasing $z$:} When we tested the four algorithms, we released the percentage threshold from 15\% to 12\%. This is because in some settings, a lot of real exceptions in the dataset have the exact of 15\% change that make them hard to detect without releasing $z$.\par
\noindent\textbf{Algorithms Evaluated:} We tested four algorithms presented in this paper: Baseline, STRATIFIED-DETECTOR, UDOMETER, and PRIORITY-UDOMETER. All these four algorithms share the same database-controlled parameter of query budget per day and the parameters $s$, $p$, and $z$ that are presented in subsection \ref{exception} that define exceptions.\par
\noindent\textbf{Performance Measures:} For estimation accuracy, we measure the percent error (i.e., $|\tilde{\theta} - \theta| / |\theta| * 100$ for an estimator $\tilde{\theta}$ of aggregates $\theta$). We also measure the precision (i.e., $|\tilde{\theta} - \theta| / |\tilde\theta| * 100$ for an estimator $\tilde{\theta}$ of aggregates $\theta$).\par

\subsection{Experimental Results}
\noindent\textbf{Recall and Precision} We start comparing the performance of all four algorithms for the recall under the default setting per-day query budget of 5000. Figure \ref{fig:recallK} shows the recall of the four algorithms when $s$ = 5000 and 0.01 $\geq p \geq$ 0.99. As shown, both PRIORITY-UDOMETER and UDOMETER improve their recall when $k$ is larger. However, when $k$ becomes very large both algorithms reach almost 100\% recall, and this is because both algorithms can almost download the entire tuples corresponding to its candidate. On the other hand, the STRATIFIED-DETECTOR and BASELINE do not show any improvement. In Figure \ref{fig:recallS}, all algorithms achieve a higher recall with bigger supported threshold. This is because when the support threshold is large, there are less candidates in the query-pool and each candidate has larger space and sample size. Figures \ref{fig:precisionK} and \ref{fig:precisionS} depicts a performance comparison between the four algorithms in terms of precision. In Figure \ref{fig:precisionK}, the impact of $k$ for both PRIORITY-UDOMETER and UDOMETER is small. And with larger $s$ the gap between these two algorithms is small. The precision for these two algorithms is not affected by the drill-downs. This is because when support threshold is large, each candidate has bigger sample size to spend. On the other hand, the gap between these two algorithms and the other two (STRATIFIED-SAMPLER and BASELINE) is very big.\par

Figure \ref{fig:recallP} shows that the change threshold does not have a big impact on the recall. And with larger $k$ and $s$, there is barely a gap between PRIORITY-UDOMETER and UDOMETER. So the change threshold nearly does not lead to any difference between these two algorithms. As for STRATIFIED-DETECTOR and BASELINE, their recall seems to improve when $p$ is large. However, and as shown in Figure \ref{fig:precisionP}, the precision for both algorithms seems to improve when $p$ is smaller.\par
In contrast, Figures \ref{fig:recallQB} and \ref{fig:precisionQB} reflect how query limitation has big impact on both recall and precision. The recall and drops from 95\% for larger query budget to close to 55\% in worst case for PRIORITY-UDOMETER  (or 35\% for UDOMETER). Moreover, the precision also drops from 95\% to 40\% for both algorithms. When $s \geq 1000$ we can only spend two query budget per-candidate to estimate its average price and judge whether it is an exception or not. Thus, we get low recall and precision. Overall, the PRIORITY-UDOMETER have higher recall and precision than UDOMETER.\par
\noindent\textbf{Score Function vs. Highest COUNT:} Figure \ref{fig:recallScoreFK}, and \ref{fig:recallScoreFS} depict the recall measurement of $S$ that used in PRIORITY-UDOMETER comparing with taking highest COUNT in UDOMETER when we choose subspace. These figures take into consideration only the candidates that contain longer conjunction (i.e., candidates with subspaces). No matter what the $k$ or support threshold are, the score function always give better prediction comparing with taking just the highest COUNT. Thus, PRIORITY-UDOMETER is more accurate for candidates with subspaces. Moreover, Figures \ref{fig:precisionScoreFK} and \ref{fig:precisionScoreFS} compare the precision under both $k$ and $s$. Similar to the recall, the PRIORITY-UDOMETER in general gives more accurate precision.\par
Finally, Figure \ref{fig:trend} depicts how PRIORITY-UDOMETER detects the overall trend for candidates, even when it incorrectly judges whether a candidate is an exception or not. The figure shows how the average price of phones with gray color drops sufficiently on Nov 20th to be an exception (i.e., which PRIORITY-UDOMETER detected correctly).\par

\section{Related Work}
\noindent\textbf{Crawling and Extraction for Hidden Databases:} There has been a number of prior work in crawling and extracting hidden databases content. This crawling requires understanding of query interfaces which was extensively studied (e.g., \cite{parsing interface, query interface}). Even though crawling the entire database (e.g., \cite{crawling}) could lead us to find all exceptions, it still unreasonable solution for this particular problem as we not only need to crawl the database one time but also for all time interval (e.g., every day!). Moreover, because of query limitations forced on hidden databases, we cannot enumerate all the selection conditions to check whether each of them is an exception or not.\par
\noindent\textbf{Aggregate Estimations over Hidden Web Databases:} Aggregate estimations over hidden web databases has been investigated over time for both static and dynamic databases. Both types use efficient techniques to obtain random samples from hidden web databases described in \cite{dynamic aggregate, static aggregate}. Unlike this paper, all prior work focuses on answering estimation with one select condition. But in this problem, estimating each select condition separately waste a lot of queries, where a sample could be used for one or more select conditions.\par
\noindent\textbf{Approximate Answers for Aggregate Queries:} Decision support applications and data mining applications execute aggregate queries over very large databases to obtain important and useful information. Existing work (e.g., \cite{stratified sampling, stratified sampling2}) introduced an efficient approach using a stratified sampling technique to optimize scalability and resources when providing approximate answers for a given workload of queries. Unlike their workload where tuples have probability of occurrence given as an input, our workload has no knowledge about the distribution function specified by the workload. Moreover, the workloads we consider do not have to be fixed nor similar and the size of these workloads could be very large, which make their optimal stratification technique unsuitable for this particular problem. Other work (e.g., \cite{join synopses}) focuses only on a specific type of query (i.e., join-queries) for approximate query answering.\par
\noindent\textbf{Data Discovery and Exploration:} Detecting anomalies on very large data, such as data cubes and data warehouses, is very important to detect problem areas or new opportunities. Most of existing work (e.g., \cite{discovery-driven, hierarchical summaries, mining outliers}) focuses on traditional multidimensional databases where a full access to these databases is a must. Other work (e.g., \cite{structured data}) discusses detecting meaningful changes over hierarchically structured data, such as nested object data, while \cite{olap} presents a single operator on OLAP products to summarize reasons for drops or increases observed at an aggregated level. Moreover, detecting changes for large scale data using sampling has been discussed (e.g., \cite{detecting changes sampling}).\par

\section{Conclusion}
In this paper, we have developed a novel technique for tracking and discovering exceptions over dynamic hidden web databases, which change over time through its restrictive web search interface. In general, our technique consists of two main phases: (1) generating a query-pool that contains all query candidates, and (2) finding all exceptions (in terms of sudden changes of aggregates) from this query-pool. We developed a stratified sampling technique along with query reissuing to guide the process of finding exceptions. We presented a comprehensive set of experiments that demonstrate the superiority of our approach over the baseline solutions on real-world datasets.


\begin{thebibliography}{9}

\bibitem{apriori}
Agrawal, R. and Srikant, R. (1994). Fast algorithms for mining association rules in large databases. \textit{Proceedings of the 20th international conference on Very Large Data Bases (VLDB'94)} (p./pp. 478--499), September, : Morgan Kaufmann. 
\bibitem{random walk}
Dasgupta, A., Das, G. and Mannila, H. (2007). A random walk approach to sampling hidden databases. \textit{Proceedings of the 2007 ACM SIGMOD international conference on Management of data} (p./pp. 629--640), New York, NY, USA: ACM. ISBN: 978-1-59593-686-8
\bibitem{dynamic aggregate}
Liu, W., Thirumuruganathan, S., Zhang, N. and Das, G. (2014). Aggregate Estimation Over Dynamic Hidden Web Databases.. PVLDB, 7, 1107-1118.
\bibitem{stratified sampling}
Chaudhuri, S., Das, G. and Narasayya, V. R. (2007). Optimized stratified sampling for approximate query processing.. \textit{ACM Trans. Database Syst.}, 32, 9.
\bibitem{parsing interface}
Zhang, Z., He, B. and Chang, K. C.-C. (2004). Understanding Web Query Interfaces: Best-Effort Parsing with Hidden Syntax. \textit{Proceedings of the 2004 ACM SIGMOD International Conference on Management of Data (SIGMOD 2004)} (p./pp. 107-118), .
\bibitem{query interface}
Kabisch, T., Dragut, E. C., Yu, C. T. and Leser, U. (2009). A Hierarchical Approach to Model Web Query Interfaces for Web Source Integration.. \textit{PVLDB}, 2, 325-336. 
\bibitem{crawling}
Raghavan, S. and Molina, H. G. (2001). Crawling the hidden web. \textit{Proceedings of the 27th International Conference on Very Large Databases (VLDB 2001)} (p./pp. 129-138), .
\bibitem{static aggregate}
Dasgupta, A., Jin, X., Jewell, B., Zhang, N. and Das, G. (2010). Unbiased estimation of size and other aggregates over hidden web databases.. In A. K. Elmagarmid and D. Agrawal (eds.), \textit{SIGMOD Conference} (p./pp. 855-866), : ACM. ISBN: 978-1-4503-0032-2
\bibitem{stratified sampling2}
Chaudhuri, S., Das, G. and Narasayya, V. R. (2001). A Robust, Optimization-Based Approach for Approximate Answering of Aggregate Queries.. In S. Mehrotra and T. K. Sellis (eds.), \textit{SIGMOD Conference} (p./pp. 295-306), : ACM. ISBN: 1-58113-332-4
\bibitem{online aggregation}
Hellerstein, J. M., Haas, P. J. and Wang, H. J. (1997). Online Aggregation.. In J. Peckham (ed.), \textit{ACMSIGMOD International Conference on Management of Data} (p./pp. 171-182), May, Tucson: ACM Press.
\bibitem{Wavelets}
Vitter, J. S. and Wang, M. (1999). Approximate Computation of Multidimensional Aggregates of Sparse Data Using Wavelets.. In A. Delis, C. Faloutsos and S. Ghandeharizadeh (eds.), \textit{SIGMOD Conference} (p./pp. 193-204), : ACM Press. ISBN: 1-58113-084-8
\bibitem{discovery-driven}
Sarawagi, S., Agrawal, R. and Megiddo, N. (1998). Discovery-Driven Exploration of OLAP Data Cubes.. In H.-J. Schek, F. Saltor, I. Ramos and G. Alonso (eds.), \textit{EDBT} (p./pp. 168-182), : Springer. ISBN: 3-540-64264-1
\bibitem{hierarchical summaries}
Agarwal, D., Barman, D., Gunopulos, D., Young, N. E., Korn, F. and Srivastava, D. (2007). Efficient and effective explanation of change in hierarchical summaries.. In P. Berkhin, R. Caruana and X. Wu (eds.), \textit{KDD} (p./pp. 6-15), : ACM. ISBN: 978-1-59593-609-7
\bibitem{mining outliers}
Ramaswamy, S., Rastogi, R. and Shim, K. (2000). Efficient algorithms for mining outliers from large data sets. \textit{SIGMOD '00: Proceedings of the 2000 ACM SIGMOD international conference on Management of data} (p./pp. 427--438), New York, NY, USA: ACM. ISBN: 1-58113-217-4
\bibitem{structured data}
Chawathe, S. S. and Garcia-Molina, H. (1997). Meaningful Change Detection in Structured Data. (p./pp. 26-37), May, Tuscon, Arizona
\bibitem{join synopses}
Acharya, S., Gibbons, P. B., Poosala, V. and Ramaswamy, S. (1999). Join Synopses for Approximate Query Answering.. In A. Delis, C. Faloutsos and S. Ghandeharizadeh (eds.), \textit{SIGMOD Conference} (p./pp. 275-286), : ACM Press. ISBN: 1-58113-084-8
\bibitem{olap}
Sarawagi, S. (1999). Explaining Differences in Multidimensional Aggregates.. In M. P. Atkinson, M. E. Orlowska, P. Valduriez, S. B. Zdonik and M. L. Brodie (eds.), \textit{VLDB} (p./pp. 42-53), : Morgan Kaufmann. ISBN: 1-55860-615-7
\bibitem{detecting changes sampling}
Cho, J. and Ntoulas, A. (2002). Effective Change Detection Using Sampling. \textit{Proceedings of the 28th International Conference on Very Large Databases (VLDB 2002)} (p./pp. 514--525), .
\end{thebibliography}
\end{document}